\documentstyle[prb,twocolumn,aps]{revtex}

\begin{document}
 \tolerance 50000

\draft

\title{Dualities in Spin Ladders } 
\author{Germ\'an Sierra$^{1}$\cite{ger} and
 Miguel A. Mart\'{\i}n-Delgado$^{2}$ } 
\address{ $^{1}$Institute of Theoretical Physics,
University of California, Santa Barbara, USA
\\ 
$^{2}$Departamento de
F\'{\i}sica Te\'orica I, Universidad Complutense. Madrid, Spain
 \\
}

\twocolumn[
\maketitle 
\widetext

\vspace*{-1.0truecm}

\begin{abstract} 
\begin{center}
 \parbox{14cm}{
We introduce a set of discrete modular transformations $T_\ell,U_\ell$ and 
$S_\ell$ 
in order to study the relationships between the different 
phases of the Heisenberg ladders obtained with all possible exchange
coupling constants. For the  2 legged ladder we show 
that the $RVB$ phase is invariant under the $S_\ell$ transformation,
while the Haldane phase is invariant under $U_\ell$.
These two phases are related by $T_\ell$.
Moreover there is a ``mixed" phase,  
that is invariant under $T_\ell$, and which under $U_\ell$ becomes
the RVB phase, while under $S_\ell$ becomes the Haldane phase. 
For odd ladders there exists only  the $T_\ell$ transformation which,
for strong coupling, maps the  effective antiferromagnetic spin 1/2 chain
into the spin 3/2 chain.  }
\end{center}
\end{abstract}

\pacs{
 \hspace{2.5cm} 
}
]
 \narrowtext

In the last two years the concept of duality has played a crucial
role in understanding non perturbative aspects
of Quantum Field Theory \cite{SW} and String Theory \cite{P}. 
Considering the traditional links between Particle Physics
and Statistical Mechanics or Condensed Matter  one
may wonder wether these latter areas could benefit
from the deeper understanding gained in the former ones.
In fact, duality
ideas have been important in the historical
development of Statistical Mechanics, as shown by the
Krammers-Wannier duality, order-disorder transformations, etc \cite{dual}. 
In this letter we shall explore the existence of 
duality symmetries in quantum spin systems defined 
on a lattice and more particularly  on arrays of coupled spin chains
known as spin ladders \cite{DR}.

Generally speaking a duality transformation is a mapping
between two models, or the same model with different parameters, 
which apparently  have different 
physical properties, but which become in a way equivalent
under the transformation. 
Dual theories usually give complementary descriptions
of the same underlying phenomena.

Let us  first stablish  what we mean by duality in a spin system.
We shall  consider the Heisenberg
Hamiltonian defined on the  $d$-dimensional 
hypercubic lattice $(d \geq 1)$,

\begin{equation}
H (\{J_{\mu} \}) = \sum_{{\bf \mu}} \;\sum_{{\bf x}} 
J_{ {\bf \mu}} \; {\bf S_x} \cdot {\bf S_{x + \mu} } 
\label{1}
\end{equation}

\noindent
where ${\bf S_x} $ 
is a spin $S$ matrix acting at the position
${\bf x} =(x_1, \dots , x_d)$, and 
${\bf \mu}_1 = (1,0, \dots,0), \dots ,
{\bf \mu}_d = (0,0, \dots, 1)$.

We shall define the dual of the Hamiltonian $(\ref{1})$
as a Hamiltonian $H_D =H( \{ J^D_{{\bf \mu}} \}) $ characterized
by a new set of coupling constants  $\{ J^D_{{\bf \mu}} \}$,
and such that the low energy spectrum 
of $H$ and $H_D$ is in one to one correspondence. 
This implies that the
free energy and the ground state energy will also be the same for
both models.

In the classical limit where the spin $S \gg 1$, the ground state
of (\ref{1}) is given by the classical minima,

\begin{equation}
{\bf S}_{{\bf x}} =  S \; {\bf n} \;\prod_{{\bf \mu}} 
\epsilon^{ {\bf x \cdot \mu}}_{{\bf \mu}}
\label{2}
\end{equation}

\noindent
where ${\bf n}$ is a 3-component unit vector and 
$\epsilon_{{\bf \mu}} = - {\rm sign} \; J_{{\mu }} $.
The energy of (\ref{2}) is given by,

\begin{equation}
E_0^{ {\rm class}} = -  S^2 \; \sum_{{\bf \mu}} \;\sum_{{\bf x}} 
|J_{ {\bf \mu}} |
\label{4}
\end{equation}

The signs of the exchange coupling constants, $\epsilon_\mu$, 
determine the type of 
order parameter which characterizes  the ground state. Thus  $\epsilon_{{\mu}}
=1 $ or  $-1$  correspond to ferromagnetic ($F$) 
or  antiferromagnetic ($A$)
order in the direction
$\mu$ of the lattice. Altogether there are $2^d$
possible classical vacua which we  denote, for $d=2$,
by the sequence,

\begin{equation} 
AA, \; AF, \; FA, \; FF
\label{5}
\end{equation}

The energies of the classical g.s. and the excitations 
are independent
of the type of vacua  (\ref{5}). 
All the classical Heisenberg models  
are equivalent. However  the quantum corrections 
drive them into  very different  quantum vacua.
Only the pure ferromagnetic system (i.e.
$J_{\mu} < 0 , \forall \mu$),   survives the quantum fluctuations,
but the non-ferromagnetic systems change deeply their structure.
The purpose of this letter is to show the
relations existing between the different  vacua 
by means of a certain type of duality transformations.

At this point it is useful to make an analogy
between 2d  spin systems and 
fermions  living on a 2d torus \cite{G}. To define 
a fermion   on a torus one has to specify the 
boundary conditions along the $a$ and $b$ cycles.  
They can be periodic ($P$) or  antiperiodic ($A$).
This gives rise to 4 possible spin structures,
labeled as $AA, AP, PA$ and $PP$, which mix under the action
of the modular transformations $T,U$ and $S$ as
follows \cite{G},

\begin{eqnarray}
& T: \; AA \leftrightarrow AP, \; PA , \; PP & \nonumber \\
& U: \; AA \leftrightarrow PA, \; AP , \; PP & \label{6} \\
& S: \; AP \leftrightarrow PA, \; AA , \; PP & \nonumber 
\end{eqnarray}

Observe that the spin structure $PP$ is left invariant under
the action of  the modular group. The fermion determinant with the
boundary conditions $AA, AP, PA$ 
turns out to be given by Jacobi  ${\vartheta}$ functions, which transform
among themselves under the modular 
group as described by (\ref{6}). The 
fermion determinant for $PP$ boundary conditions is zero due to the existence
of a zero mode.

In the case of 2d spin systems the role of the cycles $a$ and $b$ is
played by the directions ${\bf \mu}_1 =(1,0)$ and  
${\bf \mu}_2 =(0,1)$.  The analog of the spin structure
is given by the (anti)ferromagnetic nature of the coupling $J_{{\bf
\mu}}$ along the directions ${\bf \mu}_{1,2}$. 
Finally,  a modular
transformation  is a redefinition of the unit cell
of the lattice. In the case of spin  ladders the above
analogies can be collected in the  
following  dictionary,

\begin{equation}
\begin{array}{rcl} 
{\rm Torus} \;{\rm Lattice} & 
\leftrightarrow & {\rm Spin} \; {\rm Ladder} \\
a-{\rm cycle} & \leftrightarrow &  {\rm legs} \\
b-{\rm cycle} & \leftrightarrow &  {\rm rungs} \\
{\rm Antiperiodic}\; {\rm BC} & \leftrightarrow & {\rm Antiferromagnetic}
\; {\rm Coup.} \\ 
{\rm Periodic}\; {\rm BC} & \leftrightarrow & {\rm Ferromagnetic}
\; {\rm Coup.} \\ 
{\rm Modular} \;{\rm Transf.} & \leftrightarrow & 
{\rm Bond} \; {\rm Moving } \; {\rm Transf.} \end{array}
\label{7}
\end{equation}

These correspondences 
have an analogue in $d >2$. 
On a 2 legged ladder, 2-ladder for short,  
we shall define three transformations $T_\ell, U_\ell$ and
$S_\ell$ as follows. The $T_\ell$ transformation consists in the shift by one 
lattice spacing of one leg respect to the other one (see fig.1). 
The $U_\ell$ transformation consist in the permutation of the two
sites of the even rungs, while leaving invariant the odd ones (see fig.2).
Finally the $S_\ell$ transformation is defined by the equation 
$S_\ell= T_\ell U_\ell T_\ell$, and has the effect of converting all the 
vertical bonds (rungs) into horizontal ones (legs), while half
of the horizontal bonds become vertical bonds and the other half
become diagonal bonds of length $\sqrt{5}$. We remark that 
$T_\ell, U_\ell$ and $S_\ell$ do not generate the standard modular group.

Using these definitions one can see that the classical vacua
$AA, AF,FA$ and $FF$, get mixed under 
the action of $T_\ell, U_\ell ,S_\ell$  in the form described by (\ref{6}),
with the replacement: 
(anti)periodic $\leftrightarrow$ (anti)ferromagnetic.
The term ``bond moving" 
in (\ref{7}) refers to a transformation introduced by Migdal
and Kadanoff in the study of the Ising model with RG methods
\cite{KM}.   Just like a fermion on a lattice with non periodic
BC's can be regarded as
essentially the same object, we conjecture  that 
the 2-ladder, and more generally the
ladders with an even number of legs, with ``magnetic structures", 
$AA, AF, FA$, belong to the same 
universality class and are related through dual transformations.
This conjecture
implies in particular the equivalence between the RVB state
($AA$ couplings) 
and the Haldane state ($AF$ couplings) of the 2-ladder studied by 
various authors \cite{H}, \cite{W}, 
but it also suggests new equivalences 
which have not been studied so far.
We shall confine ourselves in  this letter to the case 
of the spin 1/2 ladder with 2 legs,  
trying to prove the above conjecture using 
perturbative and field theoretical techniques.
At the end we shall briefly consider the case of odd ladders.

Let us start with a toy ladder.

{\bf A $2 \times 2$ cluster}

The simplest 2-ladder has 4 spins coupled by the Hamiltonian,

\begin{equation}
H= J_a ( {\bf S}_1 \cdot {\bf S}_2 + {\bf S}_3 \cdot {\bf S}_4 ) 
+ J_b ( {\bf S}_1 \cdot {\bf S}_4 + {\bf S}_3 \cdot {\bf S}_2 ) 
\label{8}
\end{equation}

The ground state of the non-ferromagnetic Hamiltonians 
(i.e. $\epsilon_{a,b} \neq 1$)
is a singlet and therefore can be written as the 
linear combination

\begin{eqnarray}
& |\psi \rangle = \tau \,\, |a\rangle +\,\,  |b \rangle & \label{9} \\
& |a \rangle = (12)(34), \;\; |b\rangle = (14)(32) & \nonumber
\end{eqnarray}

\noindent
where $(ij)$ denotes the 
singlet valence bond state constructed out from the
spin 1/2's located at the sites $i$ and $j$. 
The transformations $S_\ell, T_\ell$ and $U_\ell$
of figure 1 become for the toy ladder elementary transpositions,

\begin{equation}
T_\ell :  3 \leftrightarrow 4,  \; 
U_\ell :  2 \leftrightarrow 3,  \;
S_\ell :  2 \leftrightarrow 4 
\label{10}
\end{equation}

\noindent
The action of $T_\ell, U_\ell, S_\ell$ on the states (\ref{9}) can be easily 
derived from (\ref{9}) and ({\ref{10}). They are given in table 1.

\begin{center}
\begin{tabular}{|c|c|c|c|}
\hline
& $T_\ell$ & $U_\ell$  &  $S_\ell$ \\
\hline
$|a \rangle$  &  - $|a \rangle $ & $ |a \rangle - |b \rangle $& 
$ \, | b \rangle $   \\
$ |b \rangle$  &-$|a \rangle + |b \rangle$ & $  - |b \rangle $& 
$ \, | a \rangle $   \\ 
\hline
\end{tabular}
\end{center}
\begin{center}
Table 1.
\end{center}

The ``modular transformations" induced on the ``modular parameter"
$\tau$ that follow from table 1 are,

\begin{equation}
T_\ell : \tau \rightarrow -( \tau + 1), \;
U_\ell : \tau \rightarrow - \tau/  ( \tau + 1), \;
S_\ell : \tau \rightarrow 1/ \tau  
\label{11}
\end{equation}

\noindent
which are similar but not identical to the standard modular
transformations of the torus. 

The ground state energy   of (\ref{8})  is,

\begin{equation}
E= - \frac{1}{2} ( J_a + J_b) - 
\sqrt{ J_a^2 + J_b^2 - J_a J_b}
\label{12}
\end{equation}

\noindent
corresponding to a value of  $\tau $ 
given by,

\begin{equation}
\tau = - 1 + \frac{J_a}{J_b} - \epsilon_b 
\sqrt{ 1 - \frac{J_a}{J_b}+ \left( \frac{J_a}{J_b} \right)^2 }
\label{13} 
\end{equation}

The values of $\tau$ obtained by
changing the  signs and strengths 
of the exchange coupling constants $J_{a,b}$, cover
the whole real axis as described in table 2.

\begin{center}
\begin{tabular}{|c|c|c|c|c|}
\hline
$(\epsilon_a,\epsilon_b)$ & $AF$ & $FF$ & $FA$ & $AA$  \\
\hline
$\tau$&
$(- \infty, -2)$  & $(-2, -1/2)$&  $( -\frac{1}{2},0)$ & $(0, \infty)$ \\
\hline
\end{tabular}
\end{center}
\begin{center}
Table 2.
\end{center}

We have included the case $FF$ which corresponds to an excited
state, since the g.s. is a spin 2 multiplet.
The $S_\ell$ transformation (\ref{11}) leaves invariant the
$AA$ and $FF$  domains, while interchanges the regions $AF$ and $FA$.
$S_\ell$-duality is an 
exact symmetry of the Hamiltonian (\ref{8}). Actually,
$\tau =1$ is a fixed point of $S_\ell$.
The $T_\ell$ and $U_\ell$ transformations are approximate symmetries
in the sense that the Hamiltonian (\ref{8}) is not mapped into a 
similar one with a redefinition of $J_{a,b}$.
However one can see that $\tau = -1/2$ is a fixed point of $T_\ell$,
while the $AA$ region $\tau >1$ is mapped under $T_\ell$ into the
$AF$ region $ \tau < -2$. Similarly $\tau = -2$ is a fixed point of
$U_\ell$, while the $AA$ region $ 0 < \tau < 1$ is mapped under $U_\ell$ into the
$FA$ region $ -1/2 < \tau < 0$. 
All these shows that eqs.(\ref{6}) hold with some caveats for the 
$2 \times 2$ cluster.

Within each domain, $AA, AF$ and $FA$, we
shall   choose   a representative state $|\tau \rangle$ with
the property of being invariant under one of the dual transformations.
The state  $ \tau=1$  can be called
a $RVB$ state since it describes  the resonance 
between two vertical and horizontal bonds. The state $\tau = -2 $
is  a  Haldane like  state ($HAL$) in the sense that it is
obtained upon forming the spin 1 state along the rungs, which 
then couple to form a singlet. Finally $\tau = -1/2 $ 
is a mixed state ($MIX$), corresponding to 
ferromagnetic chains coupled antiferromagnetically.
Moreover, each of the states 
$|RVB \rangle, |HAL \rangle$ and 
$|MIX \rangle $  gets transformed into one another
by the action of $T_\ell,U_\ell,S_\ell$.

The results are summarized in table 3.

\begin{center}
\begin{tabular}{|c|c|c|c|c|c|}
\hline
$(J_a, J_b)$   & State & $\tau$ & $S_\ell$ & $U_\ell$ & $T_\ell$  \\
\hline
$AA$ & $RVB$ & $ 1$ & $RVB$ &  $MIX$ & $HAL$    \\
$AF$ & $HAL$ & $-2$ &   $MIX$ & $HAL$ & $RVB$ \\
$FA$ & $MIX$ & $-1/2$ & $HAL$  & $RVB$ & $MIX$ \\
\hline
\end{tabular}
\end{center}
\begin{center}
Table 3. 
\end{center}

For 2-ladders with a  large number of rungs we can still
make sense of the transformation properties collected in table 3.
In that case $|RVB \rangle$ denotes the ground state of a 
ladder with $AA$ couplings, etc.

The rest of the letter will be devoted to show the validity
of table 3.

{\bf The Weak Coupling Regime: $T_\ell$-duality }

If the two legs are
weakly coupled (i.e. $ |J_a/ J_b| >>1$),
the  $T_\ell$-duality becomes  a
manifest symmetry   of the effective low energy theory.

For $J_a > 0 $  
we can use bosonization techniques to show
this fact. Indeed the effective ladder Hamiltonian 
can be written in the bosonized model  as \cite{TS},


\[
H = H_{\rm WZW} + \lambda_1 ({\bf J}_L \cdot {\bf J}_R +
\hat {{\bf J}}_L \cdot \hat{{\bf J}}_R ) 
\]
\begin{equation}
+  \lambda_2 ({\bf J}_L \cdot \hat{{\bf J}}_R +
\hat {{\bf J}}_L \cdot {{\bf J}}_R )  
 + \lambda_3   {\rm Tr} ( g {\bf \sigma} )
{\rm Tr} (\hat{g} {\bf \sigma} )
+ \lambda_4 {\rm Tr} g  {\rm Tr} \hat{g} \label{15} 
\end{equation}

\noindent
where $g$ and ${\bf J}$ 
(resp. $\hat{g}$ and ${\bf \hat{J}})$ are  the WZW field and 
current which  bosonize the upper (lower) spin-chains of the ladder.
The initial values of the different coupling constants appearing in
(\ref{15})  are given by

\begin{equation}
\lambda_1 < 0, \,\, \lambda_2 =\lambda_3 = J_b ,\,\, 
\lambda_4 =0
\label{16}
\end{equation}

\noindent

In the bosonized representation the translation of a
chain by one site is equivalent to the discrete symmetry \cite{HA},
 $g \rightarrow - g$.
Therefore the operator $T_\ell$ is realized in the WZW model by the map,

\begin{equation}
T_\ell : (g, \hat{g})  \rightarrow (- g, \hat{g}) 
\label{17}
\end{equation}

\noindent implying  that $T_\ell$ is
equivalent to
the following change of couplings,

\begin{equation}
\begin{array}{lcr}
 J_a & \stackrel{T_\ell}{\rightarrow}& J_a  \\ 
J_b & \stackrel{T_\ell}{\rightarrow}& - J_b \end{array}
\label{18} 
\end{equation}

\noindent 
Eq.(\ref{18}) illustrates the relations $T_\ell \,|RVB \rangle
= | HAL \rangle$ and  $T_\ell \,|HAL \rangle
= | RVB \rangle$, which stablish the equivalence between the
RVB and Haldane states in the weak coupling limit. 
White has observed \cite{W}
that the spins located in diagonal positions of the
2-ladder tend to form
effective spins 1, and  using,  what  we call 
 the $T_\ell$-transformation, 
he  shows the equivalence between the two phases.
This is done in \cite{W} by introducing diagonal
couplings in order to connect continously 
the  $|RVB \rangle $ and $|HAL \rangle$ states. 
What we show in this letter
is that this connection can be also thought
as a discrete modular transformation by which 
the properties of both models can be put in one to one 
correspondence.

If $J_a < 0$ both  legs are  in  a ferromagnetic state
with total spin $S_{\rm tot} = N/2$. A weak antiferromagnetic
coupling, $J_b >0$, splits this degeneracy giving a state
which, to first order in perturbation theory, is given  by the
singlet appearing in the Clebsch-Gordan  decomposition
$S_{\rm tot} \times S_{\rm tot}$. Obviously,  the latter
state is invariant under a shift of one of the legs. 
Thus the state $|MIX \rangle$, in the weak coupling
regime, is  invariant under $T_\ell-$duality, according to table 3.

{\bf The Strong Coupling Regime: $U_\ell$-duality }

In the strong coupling regime (i.e.  $|J_a/J_b| \ll1$)
the rung Hamiltonian yields
the zero order approximation, while  the leg Hamiltonian
acts as a perturbation.

The rungs,  in  a $AF$ ladder, are mostly in the spin 1 state   
which couple antiferromagnetically along the leg direction, yielding
effectively a Haldane chain.
The $U_\ell$ transformation, which simply permutes the two spins on the
even rungs, leaves invariant the corresponding 
Haldane state (i.e. $U_\ell \,\,|HAL \rangle = | HAL \rangle$).

Next we shall show using perturbation theory 
that the $RVB$ and $MIX$ states are exchanged by $U_\ell$-duality.

For  $J_b >> 1$ the rungs are in a singlet state. The g.s. energy
computed to  $2^{\rm nd}$ order in $J_a$ is given by \cite{RTR},

\begin{equation}
E_0/N = - \frac{3}{4} J_b - \frac{3}{8} \frac{J_a^2}{J_b}
\label{19}
\end{equation}

\noindent
The first excited states form a band of spin 1  magnons,

\begin{equation}
|k \rangle = \frac{1}{ \sqrt{N}} \sum_{x=1}^{N} {\rm e}^{i k x } |x \rangle
\label{20}
\end{equation}

\noindent where $|x \rangle $ denotes the state with singlets
on all rungs except at the position $x$ where it is a triplet.
The  dispersion relation $\omega(k)$ of (\ref{20}) is given,  to
$2^{\rm nd}$ order in $J_a$, by \cite{RTR},

\begin{equation}
\omega(k) = J_b + J_a \; {\rm cos} k - \frac{1}{4} 
\; \frac{J_a^2}{J_b} \; ( 3 - {\rm cos} 2k) 
\label{21}
\end{equation}

The action of $U_\ell$ on the magnons (\ref{20}) is,

\begin{equation}
U_\ell \; |k \rangle  = |k + \pi \rangle 
\label{22}
\end{equation}

Hence the spectrum of the  $RVB$ and $MIX$ states,
up to second order in $J_a$,
are exchanged
under $U_\ell$-duality, as can be seen from the following identities
satisfied by (\ref{19}) and (\ref{21}),

\begin{equation} 
\begin{array}{cc}
E_0(J_a, J_b ) = E_0(- J_a,J_b) \\
\omega(k, J_a, J_b) = \omega(k + \pi, -J_a, J_b) \end{array}
\label{23}
\end{equation}

{\bf  The Intermediate Coupling Regime: $S_\ell$-duality }

When $|J_a/J_b| \sim 1$, 
the effective theory can be obtained by mapping
the ladder into the non linear sigma model (NLSM) \cite{S,Ger,Kol}. 
The  values of  the NLSM  coupling constants are given in table 4,

\begin{center}
\begin{tabular}{|c|c|c|c|}
\hline
&  $AA$ & $AF$ &    $FA$     \\
\hline
$\theta$ & 0 & $4 \pi S $ & 0 \\
\hline
$g$ & $ \frac{1}{S} \left( 1 + \frac{J_b}{2J_a} \right)^{1/2} $ &
$\frac{1}{S }$ & $\frac{1}{S } \left(  \frac{J_b}{2|J_a|} \right)^{1/2} $ \\
\hline
\end{tabular}
\end{center}
\begin{center}
Table 4: Parameters of the 2-ladder with spin $S$.
\end{center}

From these eqs. we get the curious  relation,

\begin{equation}
g_{AA}^2 = g_{AF}^2 \;+ \; g_{FA}^2
\label{38}
\end{equation}

$S_\ell$-duality corresponds to the permutation 
of vertical and horizontal bonds. Since on a
2-ladder there are twice as much horizontal bonds
than vertical ones we expect a perfect balance between both
couplings whenever $2 |J_a| = |J_b| $. 
In this case $g_{AF}= g_{FA}= g_{AA}/\sqrt{2}$. 
The change of $\theta $ by $4 \pi S$, when going from
$FA$ to $AF$,  
does not affect the physics
of the problem and  recalls what happens
with duality transformation in field theories \cite{SW}.

Extrapolating  the  NLSM-map away the intermediate couplings
we still find an agreement with table 3.
In the strong coupling regime both $g_{AA}$ and $g_{FA}$
go to the same asymptotical value, which agrees with the fact that
the $U_\ell$ operation maps one ground state into the other. On the
other hand, the value $g_{AF}= 1/S =2 $ corresponds to the 
NSLM coupling of a spin chain with spin 1.

In the weak coupling regime the NLSM map 
is not reliable,
however we see from table 4, that in that  limit
$g_{AA} = g_{AF} \gg g_{FA}  $ which agrees with the fact that
$T_\ell$ interchages the $RVB$ and $HAL$ states. Of course in this
limit we have two weakly coupled chains,
which should  be treated with bosonization techniques.

 Summarizing our results
we can say that
the Haldane and the mixed phases  are $S_\ell-$dual, while
the RVB phase is self-dual under a $S_\ell$-transformation.

{\bf Bond Moving Dualities}

What is the origin of the duality properties of
ladders?
In Conformal Field Theory or String Theory duality 
is an expression of modular invariance. 
Something of this sort exists also  for spin systems. 
To show this we shall use
a generalization of the Migdal-Kadanoff transformations, which
consist in the substitution of  couplings between NN sites 
by other NN or NNN couplings \cite{KM}. 
This is achieved by  adding a
potential $V$ to the Hamiltonian $H$, so that  the
new Hamiltonian $H' = H+V$ has a g. s. energy ( and  
free energy) $E'$
smaller than the g.s. energy $E$  of $H$, provided 
$ \langle V \rangle = 0 $, where the v.e.v  is 
taken respect to the g.s. of
$H$ \cite{KM}. From fig. 1 we observe that
the $T_\ell-$transformation  
corresponds to the bond moving potential,

\begin{equation}
V_{T_\ell} = \sum_n ( J'_b \,\, {\bf S}_1(n) \cdot {\bf S}_2(n+1)  
- J_b \,\, {\bf S}_1(n) \cdot {\bf S}_2(n)  )
\label{25} 
\end{equation}

\noindent The  Hamiltonian obtained adding (\ref{25}) to
the ladder Hamiltonian $H(J_{a}, J_b)$, 
is a new Hamiltonian $H(J'_{a}, J'_b)$, where $J_{a,b}$ are 
given by

\begin{eqnarray}
J'_a = & J_a \label{26} \\ 
 J'_b = & J_b 
{ \langle   {\bf S}_1(n) \cdot {\bf S}_2(n) \rangle }/{
\langle   {\bf S}_1(n) \cdot {\bf S}_2(n+1)  \rangle } \nonumber
\end{eqnarray}

The expectation values in (\ref{26}) are computed with 
respect to the
g.s. of $H(J_a, J_b)$.Eq.(\ref{26}) implies 
sign$( J'_b) = - \epsilon_a \epsilon_b$, which indeed
corresponds to a $T_\ell-$transformation. 

Similarly  the bond moving transformation
which corresponds to $U_\ell$ gives,

\begin{eqnarray}
J'_b = & J_b \label{27} \\ 
 J'_a = & J_a 
{ \langle   {\bf S}_1(n) \cdot {\bf S}_1(n+1) \rangle }/{
\langle   {\bf S}_1(n) \cdot {\bf S}_2(n+1)  \rangle } \nonumber
\end{eqnarray}

Finally the $S_\ell$-tranformation can be derived from its definition
 $S_\ell = T_\ell U_\ell T_\ell$. For all these transformations, 
duality would amount to 
the equality $E(J_a, J_b)= E({J'}_a, {J'}_b)$. 
The variational principle underlying the
Migdal-Kadanoff transformation  only guarantees that 
$E(J_a, J_b) > E({J}'_a, {J}'_b)$, however after the 
results obtained above using perturbative and field
theoretical methods, we have good reasons to believe 
that the  replacement of inequalities by equalities
for the energies and free energies 
yields a good approximation. Further studies
are necessary to fully stablish these facts.

{\bf Beyond the 2-ladder}
Most of the results shown so far are generalizable to the 
case of even-ladders with periodic BC's along the rungs.
The $T_\ell$-transformation 
is given by the shift of one lattice space of the
even legs respect to the odd legs, so that the rungs become
zig-zag lines across the ladder. In fact, this definition
applies to  all types of ladders, even and odd,
with different BC's across the rungs. 
The $U_\ell$ transformation 
consists in the shift by one lattice space of the even rungs,
so that the legs become zig-zag lines along the ladder.
The new coupling constants, obtained upon these transformations,
are also given by the eqs.(\ref{26}) and  (\ref{27}).

For odd-ladders with open BC's along the rungs, 
there seems to be not a sensible definition of
the $U_\ell$ and $S_\ell$ transformations, as we did above  for $T_\ell$.
The odd-ladders of type $AA$ are in the same universality class
as the spin 1/2 antiferromagnetic Heisenberg chain, whose g.s.
we shall denote as $|A_{1/2} \rangle $. On the other hand 
the $n_\ell$-ladders ($n_\ell$: odd) 
with $AF$ couplings are, at least in the
strong coupling regime,  equivalent  to  spin $n_l/2$ antiferromagnetic
chain, whose  g.s. we shall denote  as  $|A_{n_\ell/2} \rangle $.    
The role of $T_\ell$ is to exchange
the $AA$ and $AF$ couplings, which implies that
$T_\ell \,\, |A_{1/2} \rangle = |A_{n_\ell/2} \rangle $.
The previous equivalence 
can be stablished for every  regime of couplings using,
as we did for the 2-ladder, the appropiated technique.
Thus for intermediate couplings, where we can use the mapping
of the ladder into the NLSM\cite{Ger}, 
we get the parameters $\theta_{AA} = \pi$ and  $\theta_{AF}
= \pi n_\ell$, which coincide modulo $2 \pi$. 
The odd-ladders with $FA$ and $FF$ couplings are equivalent
to ferromagnetic Heisenberg chains with spins $1/2$ and $n_\ell/2$
respectively. Their g.s. is invariant under $T_\ell$.

In summary we have seen that the even and odd ladders have quite
different duality properties which is of course a manifestation of the fact
that they both belong to different universality classes.

Acknowledgements: We would like to thank R. Shankar, E. Fradkin,
L. Balents and G. Mussardo for useful conversations. 
Work partially supported by 
the grants NSF PHY94-07194 and the DGES  (G.S.) and by
CICYT under  contract AEN93-0776
(M.A.M.-D.).

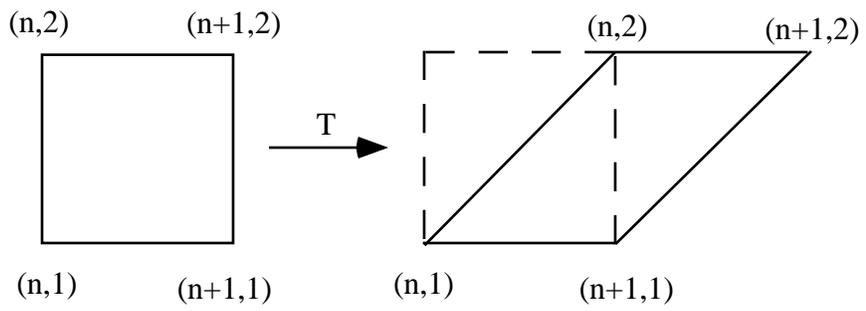
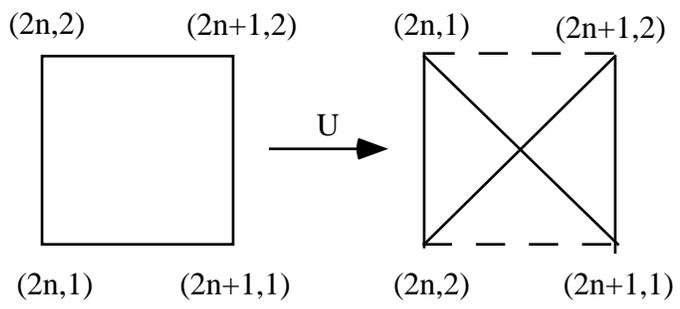
\end{document}